          [2001/04/25 1.1 (PWD)]
\documentclass[a4paper,twocolumn]{esapub} 
\usepackage{natbib}
\usepackage{graphicx}

\title{The sky behind our galaxy as seen by IBIS on INTEGRAL}
\author{L. Bassani$^1$, A. Malizia$^1$, J.B. Stephen$^1$, F. Gianotti$^1$, F. Schiavone$^1$,
A. Bazzano$^2$, A.J. Bird$^3$, L. Bouchet$^4$,T. Courvoisier$^5$, A.J. Dean$^3$, G. De Cesare$^2$, 
M. Del Santo$^2$, A. De Rosa$^2$, R. Hudec$^6$, F. Mirabel$^7$, P. Laurent$^7$, 
L. Piro$^2$, S. Shaw$^3$, A.A. Zdziarski$^8$}
\affil{$^1$ IASF-BO CNR/INAF, via Gobetti 101, 40129 Bologna, Italy}
\affil{$^2$ IASF-Ro CNR/INAF, via Fosso del Cavaliere 100, 00133 Roma, Italy} 
\affil{$^3$ School of Physics and Astronomy, Univeristy of Southampton, SO17 1BJ, UK}
\affil{$^4$ Centre d' \'Etude Spatiale des Rayonnements, CNRS/UPS, BP 4346, 31028 Toulouse France}
\affil{$^5$ INTEGRAL Science Data Center, Chemin d'\'Ecogia 16, 1290 Versoix, Switzerland}
\affil{$^6$ Astronomical Institute, Acadamy of Sciences of Czech Republic, CZ-251 65 Ondrejov, Czech Republic}
\affil{$^7$ CEA Saclay, DSM/DAPNIA/SAp, 91191 Gif-sur-Yvette Cedex, France}
\affil{$^8$ Centrum Astronomiczne im M. Kopernika, Bartycka 18, 00-716 Warszawa, Poland}

\begin{document}


\maketitle

\begin{abstract}
During the Core Programme, {\it INTEGRAL}  has surveyed a large portion of the
sky (around 9000 square degrees); although {\it INTEGRAL} is not optimized for
extra-galactic studies its observations have nevertheless given us  the
opportunity to explore the sky behind our Galaxy, something
which is impossible in some wavebands due to the presence of strong
absorption. Preliminary results from this exploration are presented and
compared with the pre-launch expectations. In particular, we detail
all extragalactic detections obtained so far: 10 active galaxies and one
cluster of galaxies. Of these sources, many have previously been studied at
energies above 10-20 keV while a few are new hard X-ray
discoveries. Since the number of detections is smaller than estimated
on the basis of the IBIS/ISGRI sensitivity, it is likely that some of
the new ISGRI sources found in this survey are extragalactic objects; 
a few of these are likely to be AGN and are described in detail.

 \end{abstract}

\section{Introduction}
The so-called "Zone of Avoidance" refers to the area contained within
$\pm$ 10-15$^\circ$ of the disk plane of the Milky Way. Gas and dust
obscure starlight within this region and screen nearly all
background extragalactic objects from traditional optical-wavelength
surveys; in the optical, as much as 20$\%$ of the extragalactic sky is
obscured by the Galaxy.  As a consequence, the Galactic plane historically has
been neglected by extragalactic astronomers. Hard X-rays
($\ge$ 10 keV) are able to penetrate this zone thereby providing a "window"
that is virtually free of obscuration not only relative to optical wavelengths
but also partly in comparison to soft X-ray observations. Unfortunately the 
hard X-ray band is
still poorly explored and the only truly all-sky survey conducted so
far dates back to the 1980's (Levine et al. 1984). This pioneering
work, made with the {\it HEAO1}-A4 instrument yielded a catalogue of about
70 sources down to a flux level of typically 1/75 of the Crab (or 2-3
$\times$ 10$^{-10}$ erg cm$^{-2}$ s$^{-1}$) in the 13-80 keV
band. Only 7 extragalactic objects are reported in the A4 survey :
none of these objects is within 10$^\circ$ of the galactic plane and
only two (Centaurus A and the Perseus cluster ) are located below
20$^\circ$ in galactic latitude. Pointed observations by {\it BeppoSAX}/PDS
have unveiled more sources but observations were sometimes limited by
the lack of imaging capability of the instrument which is particularly
crucial in the galactic plane region. A step forward in the study of
the zone of avoidance is possible with the imager on board {\it INTEGRAL},
which allows detection with a sensitivity up to a few mCrab in the
most exposed regions (i.e. the galactic center) and provides an
angular resolution of 12' and a point source location accuracy of
2-3'.  Here we present a compendium of the extragalactic results
obtained so far within the first year of the {\it INTEGRAL} Core
Programme. \\

\begin{small}
\begin{figure*}[t]
\centering
\includegraphics[width=0.8\linewidth]{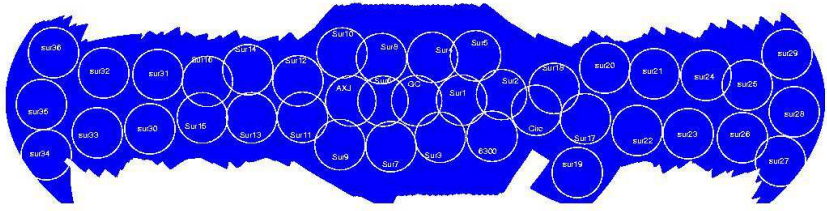}
\caption{Sky coverage so far analysed in the present survey work (-180$^{\circ}$ $<$l$<$+180$^{\circ}$, -30$^{\circ}$$<$b$<$+30$^{\circ}$). 
The circles have10$^\circ$  radius to roughly match the
half coded field of view of ISGRI.  \label{fig:single}}
\end{figure*}
\end{small}

\begin{small}
\begin{figure*}
\centering
\includegraphics[width=0.8\linewidth]{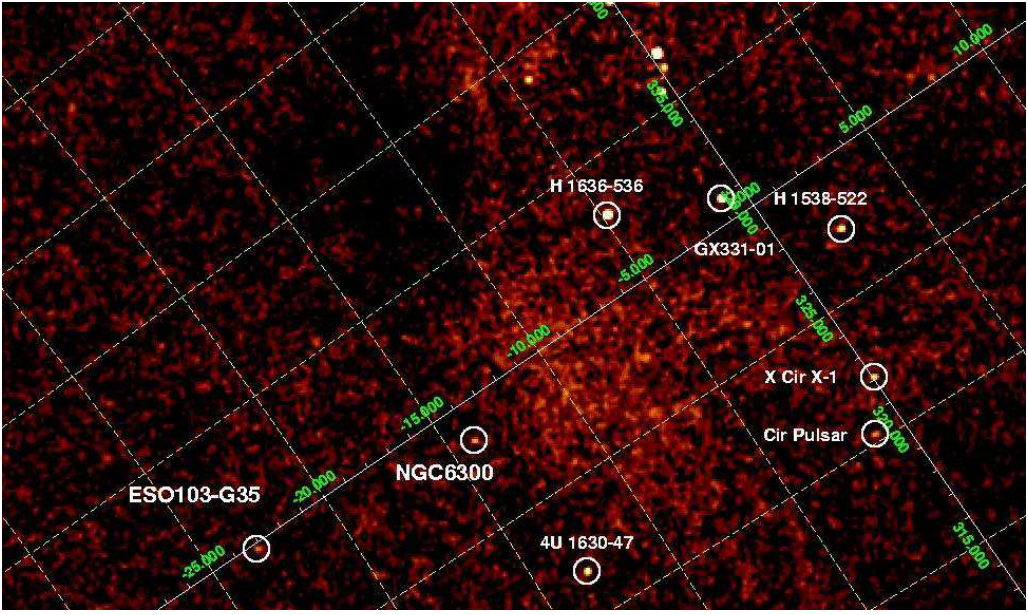}
\caption{A typical 20-100 keV ISGRI image showing 2 Seyfert 2s (ESO103-G035 and NGC6300) just above the Galactic Plane.\label{fig:single}}
\end{figure*}
\end{small}

\section{Observations and data analysis}

The IBIS coded mask instrument (Ubertini et al. 2003) on board
{\it INTEGRAL} (Winkler et al. 2003) comprises two detection
layers: ISGRI, an upper CdTe detector sensitive in the range 15 keV to
1 MeV and PICsIT, a bottom CsI detector sensitive in the range 200 keV
to 8 MeV; herein we refer only to data collected by the
first layer. The data analysed here belong to the Core Programme,
i.e. were collected as part of the {\it INTEGRAL} Galactic Plane Survey and
Galactic Centre Deep Exposure (Winkler et al. 2003) and are the just return
to the instrument teams involved in the project.  The data utilized for
this work span from orbit 20 to orbit 145 inclusive. The
total observation time varies from region to region in the sky and
ranges from 0.8 Ms in the galactic center region which is the most
exposed to typically 0.05 Ms in some parts of the galactic plane;
some regions such as those of Cygnus, Crab and Vela are more exposed
than a standard galactic plane region. During the 72 hours of the
satellite's orbital period (i.e. an {\it INTEGRAL} revolution) observations
are performed as a series of fixed pointings that last about 37 minutes
each and between which the satellite adjusts its position for the next
pointing. The data collected during one such pointing are
pre-processed and stored as one science window (scw); typically there are
110 scw in an orbit.  Data reduction was performed with the OSA 3
{\it INTEGRAL} Science Data Center analysis software (Goldwurm et al 2003),
this version of the analysis procedures and calibration files
is providing clean images although some systematic effects are still
present. Briefly, image reconstruction is performed as follows: from
the event list for one scw, a subset of events is selected according
to given energy bands. This subset is then used to build a detector
image or shadowgram from which a background map is
subtracted. Convolution of the shadowgram with a decoding array
provides the reconstructed sky image containing not only the main peak of all
sources located within the field of view but also their secondary lobes
(i.e. ghosts).  The final reconstructed sky image is obtained after
each source is identified and its secondary lobes subtracted. This
procedure is repeated for each energy band and each scw; the data of
many scw can then be mosaicked to obtain for each portion of the sky the
total exposure available.  For our search of active galactic nuclei
(AGN) we choose to search the 20-100 keV band as this provides a good
combination between sensitivity and overall background level over
the entire image: above 20 keV, most AGN have power law spectra with $\Gamma$=1.9
and a break around 100 keV (Malizia et al. 2003) so that our choice of
energy band perfectly matches these spectral characteristics.  Using the
data analysis procedure described above we sampled the whole sky by
means of 40 (mosaicked) pointings which allow the full coverage of our galaxy: see
figure 1 where circles have 10$^{\circ}$ radius to roughly match the
half coded field of view of ISGRI; note that the top and bottom parts
of the plane were also covered although not at the full instrument
sensitivity.  We then proceeded with the identification of all excesses
(above 6 sigma confidence level) visible in the images in order to
extract the detected extragalactic objects; identification of these
excesses has been performed by cross checking the ISGRI error boxes (typically
2-3 arcmin) with the Simbad/NED and HEASARC data bases. \\

\begin{table}
  \begin{center}
    \caption{Sources detected so far}\vspace{1em}
    \renewcommand{\arraystretch}{1.2}
    \begin{tabular}[h]{llll}
      \hline
      Name         & $z$      &  Type  &  $\sigma$ (scw) \\
      \hline
      GRS 1734-292  & 0.021    & Sey1   &  27 (349) \\      
      PKS 1830-21   & 2.57     &  QSO   &  16.5 (166) \\
      Circinus Gal $^{\star}$           & 0.0015   & Sey2   &  15 (23) \\
      Cygnus A                & 0.056    & NLRG   &  12 (213) \\
      Oph Cluster   & 0.028    & Cluster & 10 (298) \\
      NGC 4945 $^{\star}$               & 0.00187  & Sey2   &  9 (23)   \\
      CEN A  $^{\star}$                 & 0.0018   & NLRG/Sey2   &  8 (23) \\
      NGC 6300  $^{\star}$              & 0.0037   & Sey2   &  7.5 (43) \\
      ESO 103-G35 $^{\star}$            & 0.0133   & Sey2   &  7 (43) \\
      MCG-5-23-16$^{\star}$             & 0.00828  & Sey2   &  7 (161) \\
      NGC 6814 $^{\star}$               & 0.0052   & Sey1   &  6.6 (80) \\
      \hline \\
      \end{tabular}
Note: $^{\star}$ = objects observed by BATSE/OSSE on {\it CGRO} and PDS on {\it BeppoSAX}
    \label{tab:table}
  \end{center}
\end{table}

\begin{small}
\begin{figure}
\centering
\includegraphics[width=0.8\linewidth]{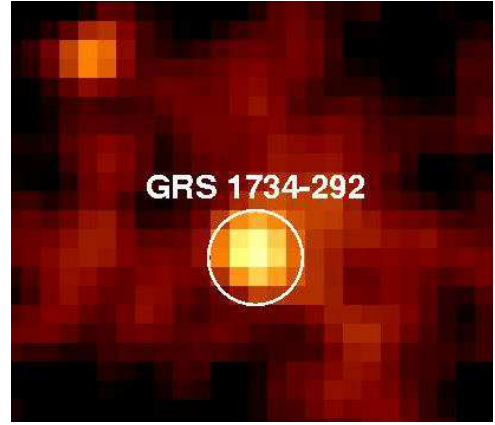}
\caption{The 20-100 keV ISGRI image of the Seyfert 1 GRS1734-292 (l=358.88$^\circ$ and
b=1.41$^\circ$). The field of view is 2$^\circ$ $\times$ 1.8$^\circ$. \label{fig:single}}
\end{figure}
\end{small}

\section{Results}
On the basis of the number of AGN seen in the {\it HEAO}-A4 survey and the
sensitivity in the energy band of interest here (typically around a few
mCrab depending on the exposure available), we have estimated
that the number of active galaxies which should be visible is around 20 when 
combining both the galactic plane survey (15) and the galactic center
deep exposure (5).  Some of these detections are expected to be known
sources already observed by previous X-ray satellites but a number
may be new hard X-ray objects. Analysis of the entire data set provides
a list of 10 AGN plus one cluster of galaxies detected by IBIS/ISGRI
in the 20-100 keV band : for each object Table 1 reports the source
name, redshift, class, number of detected sigma and science windows used in
the analysis.  Here we present some highlights of this work, while a
more detailed study of each individual source is postponed to
dedicated papers.
The first interesting result arising from table 1
is the ratio between type 2 (i.e. objects with the Broad Line Region
(BRL) hidden behind gas and dust) versus type 1 objects (i.e. those
with a visible BLR): we find a ratio of 7 to 2, which is interestingly close to
the value found in optical spectroscopic surveys (Maiolino and Rieke
(1995) and Ho et al. (1997)) if we assume that intermediate Seyfert of
type 1.8-1.9 are grouped with Seyfert 2s. 
In figure 2 a particularly rich field containing 2 Seyfert 2s (ESO103-G035 and
NGC6300) is shown, in all ISGRI images reported in this paper circles have
no physical meaning, but are shown to better high-light the detected sources.
Here too we assimilate narrow
line radio galaxies like Cen A and Cygnus A with type 2 objects while
the only QSO detected at high z is excluded from the present
evaluation. That obscured AGN must be common is obvious from the fact
that the 3 nearest (within 4 Mpc) active galaxies (NGC4945, Centaurus
A and Circinus galaxy, all detected in the present survey) are all highly obscured with N$_{H}$ $\ge$
10$^{23}$ cm$^{-2}$ (Matt et al. 2000). This situation has only
recently been appreciated due to the fact that previous observations above 10
keV of two of the above objects were not available and also perhaps due to NGC4945 being
a starburst at most wavebands and Circinus lying too close to the
galactic plane. 
Just how common obscured AGN are is still uncertain but various arguments
suggest that

\begin{small}
\begin{figure}[t]
\centering
\includegraphics[width=0.8\linewidth]{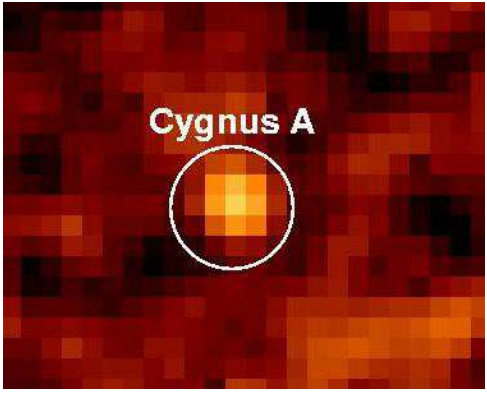}
\caption{20-100 keV ISGRI image of the Seyfert 2 Cygnus A (l=76.19$^\circ$ and
b=+5.76$^\circ$). The field of view is 1.5$^\circ$ $\times$ 2$^\circ$.\label{fig:single}}
\end{figure}
\end{small}

they are widespread, more than
typically found in optical surveys which are generally insensitive to
Compton thick sources with a high covering fraction. Hard X-rays are
better suited to uncover absorbed AGN due to their more penetrating
power; therefore our survey when completed can provide an estimate of
the real ratio of type 2 to type 1 objects.  On the other hand if
we assume that AGN have randomly oriented viewing angles,  
the ratio of type 1 to 2 objects is related to
$\Theta$ the half opening angle of
the obscuring torus, via the relation N1/N1+N2=(1-cos$\Theta$, where N1 and N2 
are the number of type 1 and 2 objects).  
From our ratio, we estimate $\Theta$ to be
around 40$^{\circ}$, in line with AGN unified model expectations.\\
Many of the sources reported in table 1 (those marked by a star)
are known hard X-ray sources and so have been previously studied in
this band by OSSE and BATSE on {\it CGRO} and/or by the PDS on {\it BeppoSAX}
(Zdziarski et al. 2000, Malizia et al. 1999, Westmore et al. 2000,
Risaliti 2002, Matt et al. 2000). Rough flux estimates based on the
detection level indicate an overall agreement with previous fluxes.
In the particular case of NGC6814, the only previous detection of
emission above 20 keV by OSSE is probably contaminated due to the
presence of a nearby ($\sim$ 30') high energy emitter (the cataclysmic
variable RXJ1940.2-1025, Bird et al. 2004), making this detection the
first "real" measurement of this Seyfert 1 in the hard X-ray domain.\\
GRS1734-292 is another source poorly known above 10 keV: it was
discovered by Granat/Art-P-Sigma in 1990 (Sunyaev et al. 1990) and
subsequently detected over a broad energy band from soft X-rays
(Barret and Grindlay 1996) to the hard X-ray band up to 400 keV
(Pavlinski et al. 1994, Churazov et al. 1992). Afterwards it was found
to be a radio jet like source and further identified with a Seyfert 1
galaxy (Marti et al. 1998). Further X-ray observations with ASCA (Sakano et al. 2002)
indicate that the source is persistent and characterized by an
absorbed power law spectrum ($\Gamma$=1.3-1.7 and N$_{H}$=10$^{22}$ 
cm$^{-2}$). The source is quite bright in the 20-100 keV band with a
flux of $\sim$ 10$^{-10}$ erg cm$^{-2}$ s$^{-1}$ (figure 3). The source lies
within the 95$\%$ error box of the unidentified EGRET source 3EG
J1736-2908; although it is a radio jet like source and so potentially
a likely counterpart of the gamma-ray source, its radio flux and
spetrum are not typical of EGRET blazars. Clearly analysis of the
ISGRI spectrum and its extrapolation to the EGRET band can shed light
on this issue.\\ Also poorly studied at high energy is Cygnus A, the
closest and best studied double radio galaxy belonging to the class of
Fanaroff-Riley type II objects.  It is interesting for a number of
reasons including the presence of radio "hot spots", its location at
the center of a cooling flow in a cluster of galaxies and the evidence
for a "buried QSO" in the nucleus. This last aspect is particularly
relevant for the ISGRI detection since gamma rays are able to penetrate heavily
absorbing material.  Cygnus A has recently been reported as a high
energy emitting object by Young et al. (2002) after detection by
RXTE-HEXTE; the overall nuclear spectrum combining high energy data
with a Chandra measurement indicate a flat power law with
$\Gamma$=1.5 and a column density of 2 10$^{23}$ cm$^{-2}$. The
source has also been observed by {\it BeppoSAX}-PDS but these data have
never been published probably in view of possible contamination from
nearby sources given its location in the galactic plane. Thanks
to the imaging capability of IBIS/ISGRI we are now able to exclude
such contamination and to analyse the PDS data which are well fitted
with a power law with $\Gamma$=1.8$\pm$0.3 and provide a 20-100 keV
flux of 7 $\times$ 10$^{-11}$ erg cm$^{-2}$ s$^{-1}$  i.e. roughly the same
brightness as seen by ISGRI (see figure 4).
At these energies the cluster contribution (kT$\sim$7) is negligible implying that 
we are detecting the radio galaxy.\\ 
At least 2 sources are new hard X-ray
detections, the Quasar PKS1830-211 and the
Oph cluster of galaxies; these are discussed below.

\begin{small}
\begin{figure}
\centering
\includegraphics[width=0.8\linewidth]{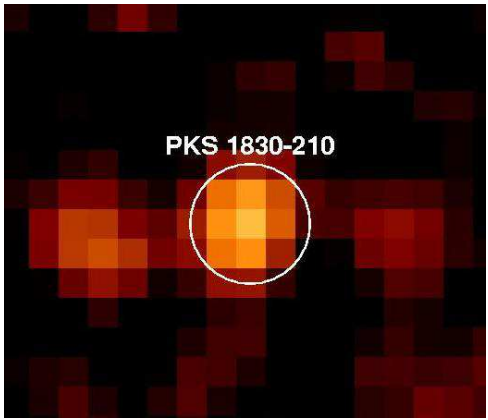}
\caption{20-100 keV ISGRI image of PKS 1830-211 (l=12.5$^\circ$ and
b=-5.70$^\circ$). The field of view is 1$^\circ$ $\times$ 1$^\circ$.\label{fig:single}}
\end{figure}
\end{small}

\begin{small}
\begin{figure}
\centering
\includegraphics[width=0.8\linewidth]{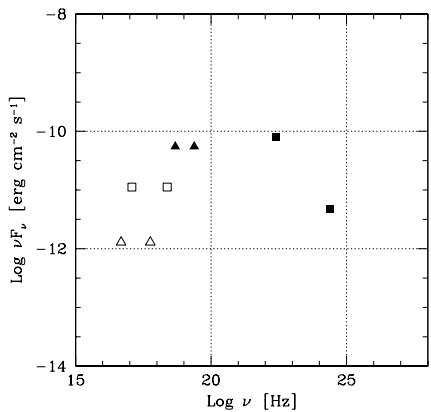}
\caption{High energy SED (soft X-rays/$\gamma$-rays) of PKS 1830211
showing a peak in the MeV region.\label{fig:single}}
\end{figure}
\end{small}

\subsection{PKS 1830-211} 

PKS 1830-211 is a quasar at a redshift of $z$=2.507. Its radio image
is gravitationally lensed by an intervening galaxy at z = 0.89
(Wiklind \& Combes 1996) into two compact images (believed to arise
from the core) separated by about 1" and a ringlike extended structure
(thought to arise from the jet) connecting the compact images (Promesh-Rao \&
Subrahmanyan 1988). The radio emission has a flat spectrum and shows a
large time variability (Lovell et al. 1998). This source is also
detected at infrared (Lidman et al. 1999), X-ray (Mathur and Nair
1997, Oshima et al. 2001), and gamma-ray (Mattox et al. 1997)
wavelengths. These detections imply that PKS 1830-211 is a blazar. PKS
1830-211 has been reported as an ISGRI source in the galactic centre
region both by Bird et al. (2004) and Revnivtsev et al. (2004): the
20-100 keV image is shown in figure 5 where this object is clearly
detected with a flux of $\sim$ 6 $\times$ 10$^{-11}$ erg cm$^{-2}$
s$^{-1}$ in the 20-100 keV band. At a redshift of 2.5 this is the
farthest object so far detected by {\it INTEGRAL}.  The broad band high
energy spectrum of the source clearly confirms its blazar nature and
further identifies it as a low frequency (i.e. MeV) peaked or red
blazar (see figure 6).

\begin{small}
\begin{figure}
\centering
\includegraphics[width=0.8\linewidth]{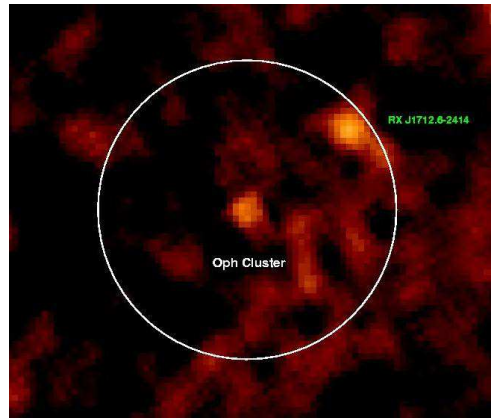}
\caption{20-100 keV ISGRI image of Oph cluster (l=0.56$^\circ$ and b=+9.27$^\circ$) and RX J1712.6-2414
(l=359.87$^\circ$ and b=+8.74$^\circ$).
In this  case the circle of 1$^\circ$ of radius, represents the PDS field of view.\label{fig:single}}
\end{figure}
\end{small}

\subsection{Oph Cluster}
The Oph cluster is a nearby ($z$=0.028) rich cluster (150 members) of
galaxies with an angular extent of roughly 1 degree.  With a 0.1-2.4 keV
luminosity of 3.1 $\times$ 10$^{-10}$ erg cm$^{-2}$ s$^{-1}$ it is the
second brightest cluster in the zone of avoidance after Perseus
(Ebelling, Mullis and Tully 2002).  The X-ray spectral data from ASCA
indicate a temperature in the range 8-26 keV (depending on the cluster
region) and an iron abundance of 0.2-0.5 relative to the cosmic value
(Watanabe et al. 2001). This source too has been reported as a
galactic centre source by Bird et al. (2004) and Revnivtsev et
al. (2004). The source is close to RXJ1712.6-2414 (a LMXB), also detected by
ISGRI, so that it is likely that a previous detection by {\it BeppoSAX}/PDS
in the high energy part of the spectrum is contaminated. Our 20-100
keV image showing the two sources and the PDS field of view can be
found in figure 7. The contaminating source is at the border of the
PDS field of view so that its contamination is probably not
significant (likely less than 10-20$\%$). This is confirmed by the
good agreement in flux between ISGRI and PDS both around 5 $\times$
10$^{-11}$ erg cm$^{-2}$ s$^{-1}$.  The PDS data provide an average
kT$\sim$ 8 keV.

\begin{small}
\begin{figure}
\centering
\includegraphics[width=0.8\linewidth]{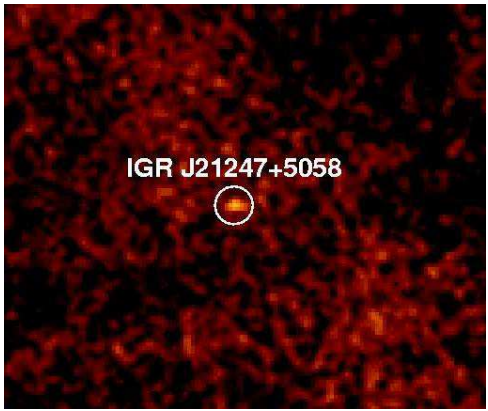}
\caption{20-100 keV ISGRI image of the unidentified sounce IGR J21247+5058
(l=93.22$^\circ$ and b=+0.38$^\circ$). The field of view is 7$^\circ$ $\times$ 6$^\circ$. \label{fig:single}}
\end{figure}
\end{small}

\section{Missing sources}
 
A comparison between our findings and expectations indicate that we are
still missing approximately half of the AGN.  In this sense 
the list of ISGRI objects not yet firmly identified deserves special attention:
around 30 objects detected in the first year of data have no obvious
counterparts in optical or infrared (Bird et al. 2004).  Most of these
objects are believed to be X-ray binary systems, where one of the two
members is either a black hole or a neutron star.  There is however
the possibility that some of them could be AGN similar to those
reported here in Table 1.  In fact, one of these new {\it INTEGRAL} objects,
IGR J21247+5058 (Walter et al. 2004) shown in figure 8, has recently been associated with the
radio source 4C50.55 by Ribo et al. (2004). These authors report that
this source has a morphology typical of a radio galaxy at 1.4 GHz: a
flat spectrum core with peak flux density of 237 mJy/beam and two
large lobes with peak flux densities of 288 and 92 mJy/beam
respectively; the lobes have spectra compatible with optically thin
synchrotron radiation.  It is also an infrared/optical source with
R-K=3.73 indicating a red spectrum possibly due to strong extinction
in the source direction.  No bright or faint ROSAT source is reported
at the position of this ISGRI object neither can we find a counterpart in
the Uhuru or A1 all sky survey catalogues, possibly suggesting that
strong absorption prevents X-rays below a few keV from escaping;
otherwise we must assume that the source is extremely
variable. Our data indicates a very bright and hard high energy
source with a 20-100 keV flux of about 10$^{-10}$ erg cm$^{-2}$
s$^{-1}$ and a 40-100 keV/20-40 keV ratio of 1.25 (Bird et al. 2004);
although a blazar type object could fulfill some of the above
characteristcs still no catalogued EGRET source is found near IGR
J21247+5058.\\ 
Another likely candidate is IGR J18027-1455 (see figure 9): inside the
2 arcmin error box of this source we have found an extended 2-MASS
infrared source (2MASXi J1802473-145454) which is also a radio object
with a flux of 10.5 mJy at 20 cm (NVSS180247-145451). The
near-infrared source (see figure  10) is fairly bright (1.44 and 1.79
mJy in J and H respectively) while the optical counterpart is dim
(B=19.3);here too we find a R-K colour index of 4 again indicating
strong reddening in the source direction.  This source is also an X-ray
emitter being detected as a faint Rosat source (1RXS
J180245.5-145432), which has a hardness ratio compatible with an AGN
interpretation (Motch et al. 1998).  The association with a possible
extragalatic object derives from the fact that most objects found to
be extended in the 2-MASS extended catalogue are galaxies although
Galactic nebulae and HII regions as well as multiple stars (mostly
double stars) and faint (mostly point-like) sources with uncertain
classifications are also possible; of all these possibilities the AGN
interpretation is more in agreement with the source being a hard X-ray
emitter.\\

\begin{small}
\begin{figure}
\centering
\includegraphics[width=0.8\linewidth]{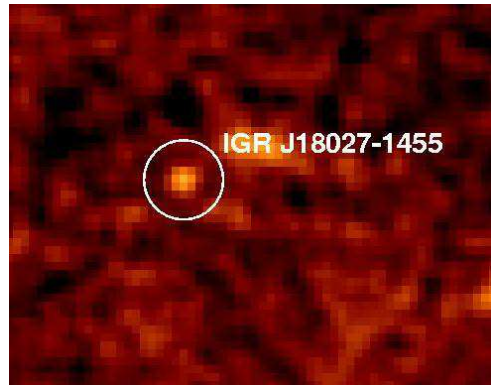}
\caption{20-100 keV ISGRI image of the unidentified sounce IGR J18027-1455
(l=14.12$^\circ$ and b=+3.68$^\circ$). The field of view is 5.3$^\circ$ $\times$ 4$^\circ$.\label{fig:single}}
\end{figure}
\end{small}

\section{Conclusions}

We have presented a compendium of the extragalactic results obtained
so far within the first year of the {\it INTEGRAL} Core Programme. Observations
of the galactic plane and center have revealed so far 11 extragalactic
objects: 10 are  active galaxies  and one is a cluster of galaxies. Of the AGN sample, 2
objects are of type 1  and 7
are of type 2. This provides a type 2 over type 1 ratio which is
in line with optical spectroscopic data and furthermore
agrees with the expectations of the unified theory. Furthermore it implies
a torus half opening angle of $\sim$40$^{\circ}$. Many of the objects
reported in this work are known to emit at high energies while a few are new hard
X-ray discoveries. In particular we find PKS 1830-211 a low
frequency (i.e. MeV) peaked or red blazar and Oph Cluster the first and only cluster
so far  reported by {\it INTEGRAL}. We also argue that a few of the new
ISGRI sources discovered during the Core Programme could also be of
extragalactic origin and describe two likely cases, IGR J21247+5058
and IGR J18027-1455.

\begin{small}
\begin{figure}
\centering
\includegraphics[width=0.8\linewidth]{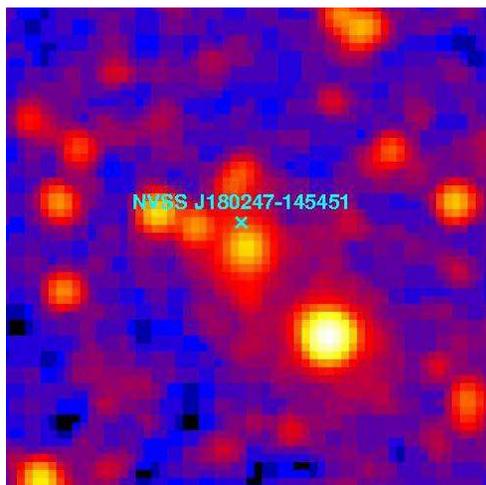}
\caption{H-band of the field (1' $\times$ 1') containing IGR J18027-1455 
showing the extended 2MASS object at the center nearby the radio source
NVSS J180247-145451.\label{fig:single}}
\end{figure}
\end{small}

\section*{Acknowledgments}
We acknowlwdge financial support by ASI (Italian Space Agency) via contract 
I/R/041/02. This research has made use of the NASA/IPAC Extragalactic Database (NED) which is operated by 
the Jet Propulsion Laboratory, California Institute of Technology, 
under contract with the National Aeronautics and Space Administration;
of the SIMBAD database, operated at CDS, Strasbourg, France;
and of data obtained from the High Energy
Astrophysics Science Archive Research Center (HEASARC), provided by NASA's
Goddard Space Flight Center.


\end{document}